\documentstyle[aps,epsfig,prb,amsmath]{revtex}

\input{prmulticol}

\begin{document}
\draft

\title{First principles elastic constants and electronic structure of
{\boldmath$\alpha$}-Pt{\boldmath$_2$}Si and PtSi} 

\author{O.~Beckstein,$^{(1)}$ J.~E.~Klepeis,$^{(2)}$ G.~L.~W.~Hart,$^{(3)}$
and O.~Pankratov$^{(1)}$}
\address{$^{(1)}$University of Erlangen-N\"urnberg, Erlangen, Germany D-91058}
\address{$^{(2)}$Lawrence Livermore National Laboratory, University of
California, Livermore, CA 94551}
\address{$^{(3)}$National Renewable Energy Laboratory, Golden CO 80401}

\date{15 March 2001}
\maketitle

\begin{abstract}
  We have carried out a first principles study of the elastic properties and
  electronic structure for two room-temperature stable Pt silicide phases,
  tetragonal $\alpha$-Pt$_2$Si and orthorhombic PtSi.  We have calculated all
  of the equilibrium structural parameters for both phases: the $a$ and $c$
  lattice constants for $\alpha$-Pt$_2$Si and the $a$, $b$, and $c$ lattice
  constants and four internal structural parameters for PtSi.  These results
  agree closely with experimental data.  We have also calculated the
  zero-pressure elastic constants, confirming prior results for pure Pt and Si
  and predicting values for the six (nine) independent, non-zero elastic
  constants of $\alpha$-Pt$_2$Si (PtSi).  These calculations include a full
  treatment of all relevant internal displacements induced by the elastic
  strains, including an explicit determination of the dimensionless internal
  displacement parameters for the three strains in $\alpha$-Pt$_2$Si for which
  they are non-zero.  We have analyzed the trends in the calculated elastic
  constants, both within a given material as well as between the two silicides
  and the pure Pt and Si phases.  The calculated electronic structure confirms
  that the two silicides are poor metals with a low density of states at the
  Fermi level, and consequently we expect that the Drude component of the
  optical absorption will be much smaller than in good metals such as pure Pt.
  This observation, combined with the topology found in the first principles
  spin-orbit split band structure, suggests that it may be important to
  include the interband contribution to the optical absorption, even in the
  infrared region.
\end{abstract}

%

%

\narrowtext

\section{Introduction}
\label{sec:introduction}

Metallic Pt silicide compounds are used to make rectifying junctions on
silicon substrates.  The Schottky barrier, which determines the activation
energy for transport of the charge carriers (holes), is 220--240 meV for
orthorhombic PtSi on p-type Si(001),\cite{Chi93,Pel90} matching an important
atmospheric ``transparency window'' in the infrared region.  For this reason
these materials are well suited to infrared detector applications.
Orthorhombic IrSi has also been used in these applications and has a Schottky
barrier of 160 meV on Si(001),\cite{Pel82} matching a lower portion of the
transparency window.  An important advantage of the silicides is that they are
more compatible with current silicon-based fabrication technology than
infrared sensitive semiconductors such as InSb and
Hg$_{\text{x}}$Cd$_{1-\text{x}}$Te.

Despite their importance in infrared detector applications, relatively little
theoretical work has been done to investigate the fundamental electronic
structure and equilibrium properties of the Pt silicides.  Bisi \emph{et
al.}\cite{Bis81} used the iterative extended Huckel method to calculate the
angular-momentum-resolved density of states (DOS) for a number of near
noble-metal silicides, including $\alpha$-Pt$_2$Si and PtSi.  Yarmoshenko
\emph{et al.}\cite{Yar97} compared x-ray photoelectron spectroscopy (XPS) and
x-ray emission spectroscopy (XES) measurements to the electronic structure
calculated using a linear muffin-tin orbital (LMTO) method for a number of 4d
and 5d silicides, including PtSi.  In order to provide a more complete
understanding of the properties of the Pt silicides we have carried out first
principles electronic structure calculations for two room-temperature stable
phases, tetragonal $\alpha$-Pt$_2$Si and orthorhombic PtSi.  In addition to
calculating all of the equilibrium structural parameters
(Sec.~\ref{sec:atomic_structure}) we have also obtained values for all of the
zero-pressure elastic constants (Sec.~\ref{sec:elastic_constants}) for both
phases.  We have investigated the electronic structure
(Sec.~\ref{sec:electronic_structure}) which is directly relevant to the
infrared detector applications of these materials.  The purpose in all of
these calculations has been to provide a fundamental understanding of the
ground state properties of the Pt silicides.  We summarize our results in
Sec.~\ref{sec:summary}.

\section{Atomic structure}
\label{sec:atomic_structure}

\subsection{Crystal structures}
\label{sec:crystal_structures}
 
Table~\ref{tab:crystal_structures} summarizes the characteristics of the
equilibrium crystal structures for each of the materials considered here.  The
pure Pt and Si constituents of the Pt silicides both crystallize in a cubic
structure under normal conditions, face-centered cubic (fcc) for
Pt\cite{Pea64} and cubic diamond for Si.\cite{LBnsIII/17a} Both cubic
structures are characterized by a single lattice constant $a$
(Table~\ref{tab:lattice_constants}).

\begin{figure}[tbp] 
  \centerline{\includegraphics[width=\linewidth]{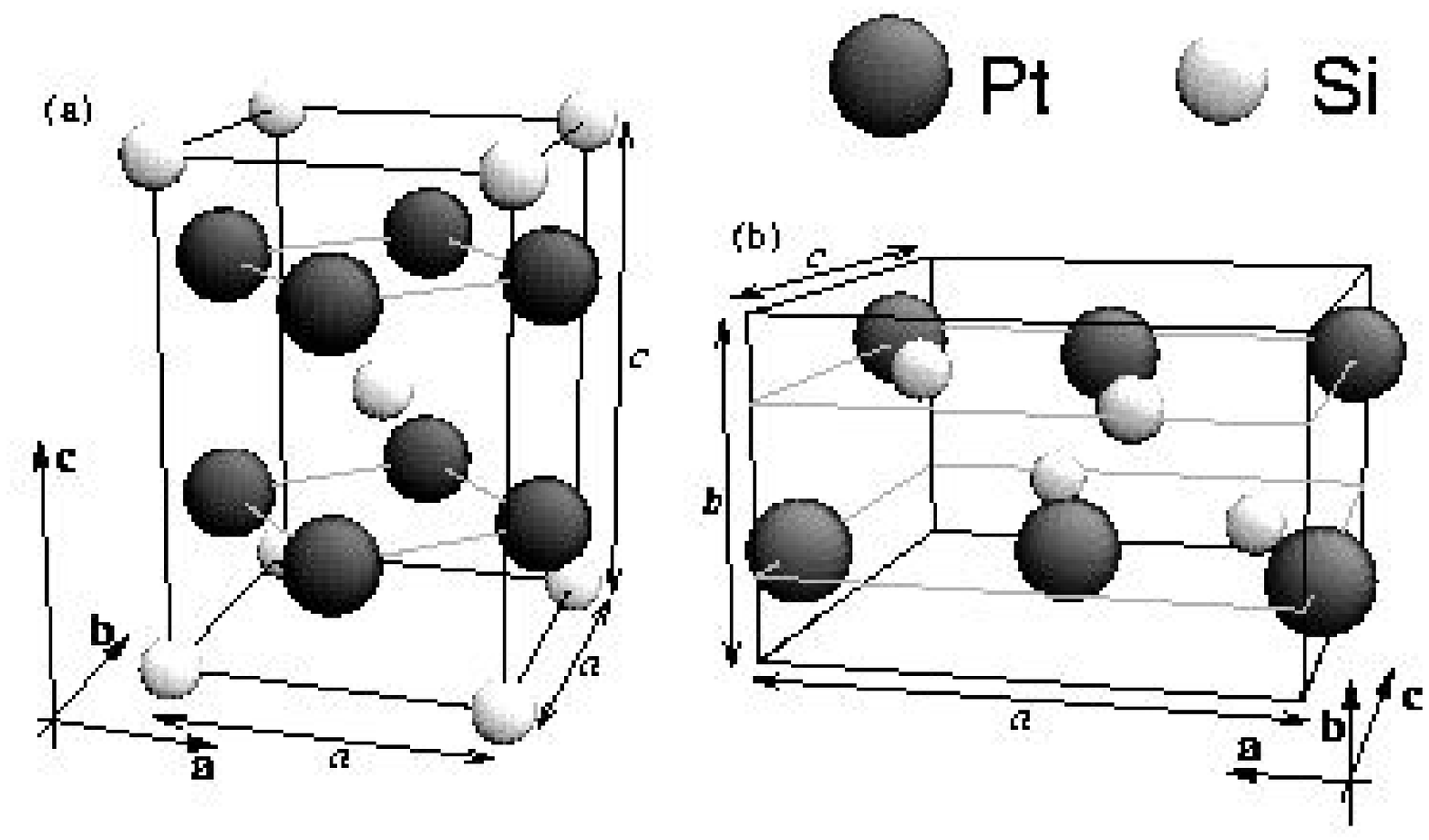}}
  \caption{Conventional unit cells of (a) body-centered tetragonal
    $\alpha$-Pt$_2$Si and (b) orthorhombic PtSi.  The relevant lattice
    constant distances are illustrated in both cases.}
  \label{fig:crystal_structures}
\end{figure}

The conventional unit cells of the two Pt silicides $\alpha$-Pt$_{2}$Si and
PtSi are
illustrated in Fig.~\ref{fig:crystal_structures}.  The structure of the
room-temperature ($T<968$~K) $\alpha$-phase of Pt$_2$Si is body-centered
tetragonal (bct) and resembles a distorted CaF$_{2}$
structure.\cite{Ram78,Goh64} A central Si atom is surrounded by eight Pt
atoms, which are located in the corners of a rectangular cell elongated along
the ${\mathbf c}$-axis.  There are two symmetry-equivalent Pt and one Si atom
in the primitive cell.  The unit cell is characterized by two lattice
constants $a$ and $c$ (Table~\ref{tab:lattice_constants}).

PtSi has a primitive orthorhombic structure [see
Fig.~\ref{fig:crystal_structures}(b)] with four symmetry-equivalent Pt and
four symmetry-equivalent Si atoms per primitive cell in an MnP-type
lattice.\cite{Pfi50,Grae73} This structure is characterized by three lattice
constants, denoted $a$, $b$, and $c$ (Table~\ref{tab:lattice_constants}).
Half of the Pt and half of the Si atoms in the primitive cell are located in a
(010) plane at ${\frac{1}{4}}b$ with the rest of the atoms in a (010) plane at
${\frac{3}{4}}b$.  The in-plane atomic coordinates are not completely specified
by the space group symmetry and thus there are four free internal structural
parameters $u_{\text{Pt}}$, $v_{\text{Pt}}$, $u_{\text{Si}}$, and
$v_{\text{Si}}$ (the ``1st position'' column in
Table~\ref{tab:crystal_structures}).  The experimental values are given in
Table~\ref{tab:internal_coordinates}.  Each Si is surrounded by six Pt atoms
at the corners of a distorted trigonal prism.  The Pt atoms have six Si
neighbors at the corners of a distorted octahedron, and four Pt neighbors,
which are positioned in four of the octahedral interstices.\cite{Grae73} The
Pt atoms are arranged in zig-zag chains along the $[100]$ direction.

\subsection{FPLMTO method}
\label{sec:nfp_method}

The equilibrium structural parameters and zero-pressure elastic constants were
calculated with a full potential linear muffin-tin orbital (FPLMTO)
method\cite{nfp97,Met99} which makes no shape approximation for the crystal
potential.  For mathematical convenience the crystal is divided into regions
inside atomic spheres, where Schr\"odinger's equation is solved numerically,
and an interstitial region.  In our FPLMTO method the basis functions in the
interstitial region are smoothed Hankel functions.\cite{Bot98} This method
does \emph{not} require the use of empty spheres, even for open structures
such as cubic-diamond-phase Si.  The atoms were treated scalar
relativistically within the local density approximation, using the
exchange-correlation potential of Ceperley and Alder.\cite{Cep80} Spin-orbit
interactions were not included.  The choice of basis functions for the Pt and
Si atoms was optimized according to the procedure described in
Ref.~\onlinecite{nfp97}.  The parameters describing the basis are listed in
Table~\ref{tab:nfp_basis} and were used in the calculations for both silicides
in addition to pure Pt and pure Si.\cite{rem:nfp_basis} The Si 3$s$, 3$p$,
3$d$, and 4$f$ as well as the Pt 6$s$, 6$p$, 5$d$, and 5$f$ were all included
as valence orbitals.  The Pt semi-core 5$s$ and 5$p$ were treated as core
orbitals and we have not used the frozen overlapped core approximation
(FOCA---see Ref.~\onlinecite{nfp97}).

The equilibrium volume $V_{0}$ and bulk modulus $B_{0}$ of Pt and Si were
determined by fitting the total energy calculated at nine different lattice
constants to a Murnaghan equation of state.\cite{Mur44,Fu83} In the case of Pt
(Si) we used a 24$\times$24$\times$24 (12$\times$12$\times$12) cubic special
$\mathbf{k}$-point mesh which gave 6912 (864) points in the full Brillouin
zone (BZ) and 182 (28) points in the irreducible wedge.  In addition, a real
space mesh is used for calculating integrals of the potential over the
interstitial region.  We used a 16$\times$16$\times$16 mesh for Pt and an
18$\times$18$\times$18 mesh for Si.  With these choices the total energies
were converged to better than 1~$\mu$Ry per atom.

In order to determine the equilibrium lattice constants of the silicides, the
total energy hypersurfaces were minimized simultaneously with respect to all
of the lattice parameters.  In the case of tetragonal $\alpha$-Pt$_2$Si,
total-energy calculations were performed at nine different values of each of
the two lattice constants $a$ and $c$ (a total of 81 calculations) in the
range $0.93 \leq a/a_{\text{exp}},\,c/c_{\text{exp}} \leq 1.07$.  For
orthorhombic PtSi we used seven different values of $a$, $b$, and $c$,
respectively (343 calculations), within the same relative ranges.  For both
$\alpha$-Pt$_2$Si and PtSi, 216 $\mathbf{k}$-points were sampled in the full
BZ which reduced to 28 and 27 special $\mathbf{k}$-points in the irreducible
wedge, respectively.  In the case of $\alpha$-Pt$_2$Si a
16$\times$16$\times$16 real space mesh was used for the interstitial integrals
and a 24$\times$16$\times$24 mesh was used for PtSi.  With these choices the
energy per atom differed from the fully converged value (which was found for
approximately 13\,000 ${\mathbf k}$-points in the full BZ) by 0.20~mRy for
$\alpha$-Pt$_2$Si and 0.19~mRy for PtSi.  The resulting total energy
hypersurfaces, $E(a,c)$ and $E(a,b,c)$, were each fit to a third order
polynomial in the lattice parameters.  These polynomials were then minimized
to yield the equilibrium parameters.  The bulk modulus and equilibrium volume
were obtained by fitting the total energy as a function of volume to a
four-term Birch-Murnaghan equation of state,\cite{Bir78}
\begin{equation}
  \label{eq:eos}
  E(V) = \sum_{n=1}^{4} a_{n} V^{-2n/3}.
\end{equation}
This fit was differentiated twice, yielding the bulk modulus
$B_{0}=V\left[\frac{\partial^{2} E(V)}{\partial V^2}\right]_{V_{0}}$ at the
theoretical equilibrium volume.

The four free internal structural parameters of orthorhombic PtSi,
($u_{\text{Pt}}$, $v_{\text{Pt}}$, $u_{\text{Si}}$ and $v_{\text{Si}}$ in
Table~\ref{tab:crystal_structures}), were determined self-consistently by
calculating the {\em ab initio\/} forces\cite{nfp97,Met93} on the ions and,
within the Born-Oppenheimer approximation,\cite{Ash76} relaxing the position
of each individual atom in the direction of the forces until the absolute
values of the forces were converged to less than 1.5 mRy/a.u. 512 special
$\mathbf{k}$-points were used within the full BZ (corresponding to 64 in the
irreducible wedge).  Initially the atomic positions were relaxed starting from
the experimental structure\cite{Grae73} and holding the three lattice
constants fixed at their experimental values.  Using these theoretically
determined internal parameters, the theoretical equilibrium lattice constants,
$a_{0}$, $b_{0}$, and $c_{0}$, as well as the bulk modulus $B_{0}$, were
determined using the procedure described above.  A second geometry relaxation
was then carried out but now holding the lattice constants fixed at these
theoretically determined equilibrium values.  This yielded a second set of
internal structural parameters which we refer to as the self-consistent
theoretical values.  In principle this cycle could be repeated many times to
obtain a set of lattice constants and internal parameters which are truly
``self-consistent.''\cite{rem:geomSC} However, in practice we find that after
the first cycle there are only small differences between the two sets of
internal parameters (see Table~\ref{tab:internal_coordinates}) and so we
regard them as being converged.  We also note that the value of $B_0$ obtained
using fixed values of the internal structural parameters, as described here,
is not strictly correct and that we relax the constraint of fixed internal
parameters when we discuss the elastic constants (including $B_{0}$) in
Sec.~\ref{sec:ptsi_elastic}.

\subsection{Equilibrium properties}
\label{sec:equilibrium}

In order to test our method\cite{nfp97} and, in particular, to test our choice
of basis functions, we calculated the equilibrium lattice constant $a_{0}$ and
bulk modulus $B_{0}$ for Pt and Si as described in Sec.~\ref{sec:nfp_method}.
The results are given in Table~\ref{tab:lattice_constants} and compared to
experimental data.  Since we will focus next on the elastic constants, we pay
particular attention to $B_{0}$, which is essentially an elastic constant.  We
see in Table~\ref{tab:lattice_constants} that we obtain rather good agreement
between our values and the experimental data.  The self-consistent equilibrium
lattice-constants $a_{0}$ and $c_{0}$ for $\alpha$-Pt$_2$Si are also listed in
Table~\ref{tab:lattice_constants}, along with the theoretical $a_{0}$,
$b_{0}$, and $c_{0}$ lattice constants for PtSi.  The bulk moduli $B_{0}$ at
the theoretical volumes for both Pt silicides are also given.

In Table~\ref{tab:internal_coordinates} we list the internal structural
parameters for PtSi.  The values calculated using the experimental lattice
constants are very close to those measured by Graeber \emph{et
al.}\cite{Grae73} Comparing the atomic positions using the experimental versus
the theoretical internal parameters, we find absolute shifts in the positions
of less than 0.028~a.u.  The self-consistent internal parameters obtained
using the theoretical lattice constants are generally even closer to the
experimental values.  In this case the absolute atomic shifts relative to the
experimental geometry are less than 0.023~a.u.

The theoretical cohesive energies $E_{\text{coh}}$ and heats of formation
$\Delta H_{\text{f}}$ for the self-consistent equilibrium atomic geometries
are compared to the experimental values in Table~\ref{tab:cohesive_energies}.
As is typically the case, our local-density-functional based calculations
overestimate the cohesive energy.  However, the calculated heats of formation
are much closer to the experiment.  The experimental heats of formation are
given for $T=298.15$~K whereas the theoretical values correspond to $0$~K and
do not include corrections for zero-point vibrations.  We note that the heats
of formation are very similar for the two silicides and that a plot of the
theoretical $\Delta H_{\text{f}}$ as a function of atomic percent Pt is
concave up, as required for the silicides to both be thermodynamically stable.

\section{Elastic constants}
\label{sec:elastic_constants}

\subsection{Method of calculation}
\label{sec:method_elastic}

The elastic constants determine the stiffness of a crystal against an
externally applied strain.  For small deformations we expect a quadratic
dependence of the crystal energy $E$ on the strain (Hooke's law).  The elastic
constants $c_{ijkl}$ describe this quadratic behavior.  Consider a
displacement ${\mathbf u}({\mathbf R})$ which takes every Bravais lattice
point ${\mathbf{R}}$ of the undistorted lattice to a new position ${\mathbf
R}'$ in the strained lattice,
\begin{equation}
  \label{eq:strained_lattice1}
  R_{i}' = R_{i} + u_{i}({\mathbf R}),
\end{equation}
where the index $i$ corresponds to Cartesian coordinates.  If we assume the
applied strain is homogeneous (uniform throughout the crystal), we can rewrite
Eq.~\ref{eq:strained_lattice1} as
\begin{equation}
  \label{eq:strained_lattice2}
  R_{i}' = \sum_{j} \alpha_{ij} R_{j} \quad \text{with} \quad \alpha_{ij}
  = \delta_{ij} + \frac{\partial u_{i}({\mathbf R})}{\partial R_j}.
\end{equation}
For a homogeneous applied strain the displacement gradients $\partial
u_{i}({\mathbf R})/\partial R_j$ are simply constants, independent of
${\mathbf R}$.  These displacement gradients define the nine components of a
tensor.  However, since the total energy $E$ cannot change under rotations of
the crystal as a whole, $E$ can only depend on the symmetric part of the
deformation,\cite{Nye72} called the strain tensor {\boldmath$\epsilon$}:
\begin{equation}
  \label{eq:def_strain}
  \epsilon_{ij} = \frac{1}{2}
  \left[\frac{\partial u_{i}({\mathbf R})}{\partial R_j}
  + \frac{\partial u_{j}({\mathbf R})}{\partial R_i}\right].
\end{equation}
Expanding the internal energy $E(V,${\boldmath$\epsilon$}$)$ of the crystal
with respect to the strain tensor gives\cite{Wal72}
\begin{equation}
  \label{eq:strain_expansion}
  \begin{split}
    E(V,\{\epsilon_{mn}\}) = E(V) &+ V \sum_{ij}\sigma_{ij} \epsilon_{ij} \\
    &+ \frac{V}{2} \sum_{ijkl} c_{ijkl} \epsilon_{ij} \epsilon_{kl} + \dots,
  \end{split}
\end{equation}
where the stress tensor {\boldmath$\sigma$} is defined by
\begin{equation}
  \label{eq:def_stress}
  \sigma_{ij}=\frac{1}{V} \left[\frac{\partial E(V,\{\epsilon_{mn}\})}
  {\partial \epsilon_{ij}}\right]_{\bbox{\epsilon}=0},
\end{equation}
the second order adiabatic elastic constants are given by
\begin{equation}
  \label{eq:def_c}
  c_{ijkl} = \frac{1}{V} \left[\frac{\partial^{2}E(V,\{\epsilon_{mn}\})}
  {\partial \epsilon_{ij} \partial \epsilon_{kl}}\right]_{\bbox{\epsilon}=0},
\end{equation}
and $V$ is the volume of the unstrained crystal.  It is convenient to use
Voigt notation which takes advantage of the symmetries of the tensors:
$xx\rightarrow 1,\ yy\rightarrow 2,\ zz\rightarrow 3,\ yz\rightarrow 4,\
xz\rightarrow 5$ and $xy\rightarrow 6$.  Using this notation
Eq.~(\ref{eq:strain_expansion}) becomes\cite{Nye72}
\begin{equation}
  \label{eq:voigt_expansion}
  E(V,\{e_{i}\}) = E(V) + V \sum_i \sigma_{i} e_{i}
                 + \frac{V}{2} \sum_{ij} c_{ij} e_{i} e_{j} + \dots
\end{equation}
with the strain tensor given by
\begin{equation}
  \renewcommand{\arraystretch}{1.5}
  \label{eq:voigt_strain}
  \bbox{\epsilon}=\left(
    \begin{array}{ccc}
      e_{1}            & \frac{1}{2}e_{6} & \frac{1}{2}e_{5}\\
      \frac{1}{2}e_{6} & e_{2}            & \frac{1}{2}e_{4}\\
      \frac{1}{2}e_{5} & \frac{1}{2}e_{4} & e_{3}
    \end{array}
    \right).
\end{equation}

In order to calculate all $M$ elastic constants of a crystal we applied $M$
independent strains {\boldmath$\epsilon$}$^{(I)}$ to the unit cell, using
Eqs.~(\ref{eq:strained_lattice2}) and (\ref{eq:def_strain}) to determine the
atom positions within the strained unit cell.  In particular, we have $M=3$
for both cubic Si and cubic Pt, $M=6$ for tetragonal Pt$_{2}$Si, and $M=9$ for
orthorhombic PtSi.  Each strain $I = 1, \dots, M$ was parameterized by a
single variable $\gamma$ and we calculated the total energy $E^{(I)}(\gamma)$
for a number of small values of $\gamma$.  For these small distortions,
$E^{(I)}(\gamma)$ was fit to a polynomial in $\gamma$ and then equated to the
appropriate elastic constant expression $E(V,\{e^{(I)}_{i}(\gamma)\})$ in
Eq.~(\ref{eq:voigt_expansion}).  From all of the fits we obtained a system of
$M$ linear equations for the elastic constants, which was solved for the
$c_{ij}$.  Since we always take the undistorted crystal to be the
zero-pressure theoretical equilibrium structure, the applied stress
{\boldmath$\sigma$} is zero and so the second term of
Eqs.~(\ref{eq:strain_expansion}) and (\ref{eq:voigt_expansion}) does not enter
in the calculations described here.

The parameterizations we used for the three independent strains in the cubic
cases of Pt and Si are given in Table~\ref{tab:cubic_strains}. Strain $I=1$ is
a volume-conserving stretch along the ${\mathbf z}$-axis, the second strain is
equivalent to simple hydrostatic pressure, and strain $I=3$ corresponds to a
volume-conserving monoclinic shear about the ${\mathbf z}$-axis.  We carried
out calculations for 9 values of $\gamma$ in the range of $-0.01$ to $0.01$
for strains 1 and 2.  However, for strain 3 we calculated 9 points in the
range from $-0.04$ to $0.04$ because the changes in the energy were rather
small (a maximum of 0.1~mRy for $\gamma=0.01$), leading to larger error
estimates in the case of the smaller range.  In order to calculate the six
independent and non-vanishing elastic constants of tetragonal
$\alpha$-Pt$_2$Si we used the strains given in
Table~\ref{tab:pt2si_strains}.\cite{rem:pt2si_strains}  Orthorhombic PtSi has
nine independent elastic constants and we chose the nine strains listed in
Table~\ref{tab:ptsi_strains}.  For each of the silicide strains we carried out
calculations for seven values of $\gamma$ in the range of $-0.01$ to $0.01$,
except for strains 8 and 9 in the case of PtSi where only five values of
$\gamma$ were considered (these monoclinic strains were particularly CPU
intensive).  Calculational errors in the elastic constants were determined
from the least-squares fit to $E(\gamma)$.  All of our results were obtained
from fits of the energy to third order in $\gamma$ because these yielded the
smallest errors compared to polynomials of order two and four; the single
exception was Pt $c_{44}$, where minimum standard errors resulted from a
fourth order fit.

Calculations of the elastic constants require a very high degree of precision
because the energy differences involved are of the order of $10$ to $1000\
\mu$Ry.  This circumstance requires the use of a fine $\mathbf k$-point mesh.
With our choice of 23\,328 special ${\mathbf k}$-points in the full BZ for Pt,
864 for Si, and 5832 for $\alpha$-Pt$_2$Si and PtSi, the energy per atom was
converged to $1\,\mu$Ry or better in all cases.  In order to minimize
numerical uncertainties we used the same $\mathbf k$-point mesh for all of the
calculations in a given material.  The differing symmetries of the various
strains $I$ resulted in differing numbers of irreducible $\mathbf k$-points.
We also checked that we obtained the same total energy for $\gamma=0$,
regardless of strain $I$ (and hence different symmetry and irreducible
${\mathbf k}$-points).  All of the calculations were carried out at the
theoretical equilibrium lattice constants listed in
Table~\ref{tab:lattice_constants}.  Relaxation of the internal degrees of
freedom was carried out in the case of all nine PtSi elastic constants.  These
relaxations are necessary because the atomic positions are not completely
fixed by the space group symmetry, even for the unstrained crystal, and
consequently there exist free internal parameters (see
Table~\ref{tab:internal_coordinates}) which must be redetermined for any
distortion of the crystal, including hydrostatic pressure.  Relaxations were
also carried out in those cases where the strain-induced symmetry-reduction
prompted it ($c_{44}$ for Si and strains 1, 4, 6 for $\alpha$-Pt$_2$Si).  For
comparison we have calculated ``frozen'' elastic constants in these same
cases, where the internal structural parameters where frozen at their
zero-strain equilibrium values.

\subsection{Pt}
\label{sec:pt_elastic}

The three elastic constants for Pt are listed in
Table~\ref{tab:pt+si_elastic}.  Pt is the only one of the metals considered in
this work for which experimental data on elasticity is available.  MacFarlane
\emph{et al.}\cite{Mac65} extrapolated the values to 0~K, which makes them
well suited for a comparison to our zero-temperature calculations.  In the
case of $c_{11}$ and $c_{12}$ we find good agreement between our results and
the experimental data (within 3--4\%).  The value of $c_{44}$ deviates by
14\%, although the absolute error is approximately 10~GPa for all three
elastic constants.  The error in $c_{44}$ can be understood if we look closely
at the band structure.  Pt exhibits a wealth of van Hove singularities
directly at the Fermi energy, making it difficult to integrate over the Fermi
surface.  A high density of ${\mathbf k}$-points (23\,328 in the full BZ) and
a very small smearing width of 7~mRy in the higher-order smearing
procedure\cite{Met89} are essential because the Fermi energy and hence the
total energy depend quite sensitively on these parameters.  The calculated
value of $c_{44}$ was found to be more sensitive to the ${\mathbf k}$-points
than the other two elastic constants.  Conversely, the silicides did not
warrant such a special treatment and were calculated with a smearing width of
25~mRy.  It seems plausible that a more accurate treatment of the elastic
properties of Pt may also require inclusion of spin-orbit coupling.  The bulk
modulus calculated from the theoretical values of the elastic constants
[$B_{0}=\frac{1}{3}(c_{11} + 2 c_{12})$] is 290.8~GPa.  It agrees well with
both the experimental value of 288.4~GPa and the one extracted from the fit to
a Murnaghan equation of state, 287.8~GPa (Sec.~\ref{sec:equilibrium}).

The requirement of mechanical stability in a cubic crystal leads to the
following restrictions on the elastic constants\cite{Wal72}
\begin{equation}
  \label{eq:cubic_stability}
  (c_{11} - c_{12}) > 0, \quad c_{11} > 0 \quad c_{44} > 0,
    \quad (c_{11} + 2 c_{12}) > 0.
\end{equation}
The Pt elastic constants in Table~\ref{tab:pt+si_elastic} obey these stability
conditions, including the fact that $c_{12}$ must be smaller than $c_{11}$.
These conditions also lead to a restriction on the magnitude of $B_0$.  Since
$B_0$ is a weighted average of $c_{11}$ and $c_{12}$ and stability requires
that $c_{12}$ be smaller than $c_{11}$, we are left with the result that $B_0$
is required to be intermediate in value between $c_{11}$ and $c_{12}$,
\begin{equation}
  \label{eq:cubic_b0}
  c_{12} < B_0 < c_{11}.
\end{equation}

\subsection{Si}
\label{sec:si_elastic}

Because Si has a cubic structure, it has only three distinct, non-vanishing
elastic constants.  These were determined with the same strains as in the case
of Pt (Table~\ref{tab:cubic_strains}).  Our results are close to experiment,
as indicated in Table~\ref{tab:pt+si_elastic}.  The bulk moduli from the total
energy minimization and from the elastic constants [$B_{0}=\frac{1}{3}(c_{11}
+ 2 c_{12})$] have the same value of 95.9~GPa, close to the one calculated
from the experimental elastic constants, 97.0~GPa.  The Si elastic constants
in Table~\ref{tab:pt+si_elastic} also obey the cubic stability conditions in
Eq.~(\ref{eq:cubic_stability}), meaning that $c_{12} < B_0 < c_{11}$.

It is perhaps worth noting that the calculation of $c_{44}$ required a
relaxation of the positions of the Si atoms within the distorted unit cell.
The symmetry reduction by the monoclinic shear ({\boldmath$\epsilon$}$^{(3)}$
in Table~\ref{tab:cubic_strains}) allowed the Si atoms to relax in the $[001]$
direction.  Without this relaxation, $c_{44}$ would have been 108.6~GPa; this
is to be compared with the relaxed value of 79.9~GPa and the experimental
value of 79.1~GPa.\cite{LBnsIII/29a} We have also obtained the dimensionless
Kleinman internal displacement parameter $\zeta$ which determines the
magnitude of the internal displacements along the $[001]$ direction,
\begin{equation}
  \label{eq:si_internal_disp}
  \Delta u_{3}^{(I=3)} = \zeta {\frac{a}{4}} \epsilon_{xy},
\end{equation}
where $a$ is the lattice constant and $\epsilon_{xy} = {\frac{1}{2}}e_{6}$ is the
appropriate element of the strain tensor [Eq.~(\ref{eq:voigt_strain}) and
Table~\ref{tab:cubic_strains}].  Fitting our calculated values of $\Delta
u_{3}^{(I=3)}$ to a quadratic function in $\gamma$ we find a value of $\zeta =
0.53$ which agrees very well with the experimental value of 0.54.\cite{Cou87}

\subsection{{\boldmath$\alpha$}-Pt{\boldmath$_2$}Si}
\label{sec:pt2si_elastic}

We have applied the six strains listed in Table~\ref{tab:pt2si_strains} in
order to determine the elastic constants of tetragonal $\alpha$-Pt$_2$Si.  The
orthorhombic strains 1 and 4, and the monoclinic strain 6 all reduce the
symmetry of the crystal in such a way that the positions of the Pt atoms are
no longer completely fixed by the symmetry.  The strain-induced forces drive
them into energetically more favorable positions.  However, the Si atom
occupies a center of inversion symmetry and thus Si internal displacements are
forbidden in all cases (i.e. the strain-induced forces are identically zero).
Symmetry also places specific restrictions on the nature of the Pt
displacements.  The symmetry of strain 6, corresponding to $c_{44}$, allows Pt
internal displacements along both the $[010]$ and $[001]$ directions, while
the inversion operation leads to the requirement that the displacements must
be equal and opposite for the two Pt atoms in the primitive cell.  The
symmetry of strain 4, corresponding to $c_{11}$, is the same as the symmetry
of strain 1 and both allow internal displacements only along $[001]$.  Once
again the presence of inversion requires that the displacements of the two Pt
atoms be equal and opposite.  Strain 5, corresponding to $c_{33}$, and strain
2 result in the same symmetry as the unstrained crystal and therefore there
are no internal displacements associated with these cases, since there are no
degrees of freedom in the internal atomic coordinates of the unstrained
crystal.  Strain 3 does lower the symmetry but internal displacements are
still symmetry-forbidden.  This fact, combined with the lack of displacements
associated with strain 2 means that $c_{66}$ is unaffected.  We note that
since strains 1 and 4 result in the same symmetry reduction relative to the
unstrained crystal there will necessarily be a formal symmetry-required
relationship between the internal displacements for these two strains.  This
relationship is obtained directly from the first principles calculations.

In addition to placing restrictions on the nature of the internal
displacements, symmetry also constrains the corresponding changes in the
elastic constants themselves.  We have already seen that the values of
$c_{11}$ and $c_{44}$ are both allowed to change as a result of internal
displacements but that $c_{33}$ and $c_{66}$ must both remain unchanged.  The
bulk modulus $B_{0}$ is also required to be unchanged because it represents
the crystal response to hydrostatic pressure, corresponding to a strain
{\boldmath$\epsilon$}$^{(B)}=\gamma \delta_{ij}$ which preserves the full
symmetry of the unstrained crystal, just as in the case of strains 2 and 5.
The expression for the bulk modulus in terms of the elastic constants is
$B_{0}=\frac{1}{9}(2 c_{11} + c_{33} + 2 c_{12} + 4 c_{13})$ while the energy
expression corresponding to strain 2 is $(c_{11} + c_{12} - 4 c_{13} + 2
c_{33})\gamma^{2}$ (see Table~\ref{tab:pt2si_strains}).  Our symmetry
arguments have required that neither of these expressions can change as a
result of internal displacements and therefore the changes in $c_{11}$,
$c_{12}$, and $c_{13}$ must exactly cancel from these two expressions (we have
already shown in conjunction with strain 5 that $c_{33}$ cannot change).  The
only way to achieve both cancellations is if the displacement-induced change
in $c_{13}$ is identically zero and if the changes in $c_{11}$ and $c_{12}$
are equal and opposite.  Moreover, since $c_{11}$ appears as the sole
coefficient in the energy expression corresponding to strain 4 and since
internal displacements can only lower the energy, we conclude that the value
of $c_{11}$ must either decrease or remain the same.  This conclusion leads to
the seeming paradox that if $c_{11}$ decreases then symmetry requires that
$c_{12}$ must increase which appears to contradict the fact that internal
displacements must always lower the energy.  The resolution of this seeming
paradox comes from the fact that it is not possible to construct a strain in
which $c_{12}$ appears as the sole coefficient in the expression for the
strain energy.  It always appears in conjunction with $c_{11}$ and we have
already seen that $c_{11} + c_{12}$ is required by symmetry to be unchanged
while $c_{11} - c_{12}$ can either decrease or remain unchanged.

Our results for the six independent and non-zero elastic constants of
$\alpha$-Pt$_2$Si are given in Table~\ref{tab:pt2si_elastic}.  We have
calculated the elastic constants for the ``frozen'' configuration [all atoms
held at the positions determined solely from Eq.~(\ref{eq:strained_lattice2})]
and with the relaxation of the strain-induced forces on the Pt atoms.  In
keeping with our general symmetry arguments, we find a relaxation-induced
softening of $c_{11}$ by 4\% and of $c_{44}$ by 17\%.  In addition, $c_{12}$
increases by 6\% while the remaining elastic constants are unchanged to within
numerical uncertainties.  Our results are also consistent with the symmetry
requirement that the changes in $c_{11}$ and $c_{12}$ be equal and opposite,
since $c_{11}$ decreases by 14.8$\pm$1.5~GPa whereas $c_{12}$ increases by
14.6$\pm$1.6~GPa.  The bulk moduli calculated from the tetragonal elastic
constants and from the fit to a Birch-Murnaghan equation of state are almost
the same, giving a consistent prediction of $B_{0} = 235$~GPa.  As required by
symmetry, the bulk modulus has the same value in the frozen and relaxed
calculations.

The requirement that the crystal be stable against any homogeneous elastic
deformation places restrictions on the elastic constants, just as in the cubic
case.  For tetragonal crystals these mechanical stability restrictions are as
follows\cite{Wal72}
\begin{equation}
  \label{eq:pt2si_stability}
  \begin{split}
    (c_{11} - c_{12}) > 0, \quad (c_{11} + c_{33} - 2 c_{13}) &> 0, \\
    c_{11} > 0, \quad c_{33} > 0, \quad c_{44} > 0, \quad c_{66} &> 0, \\
    (2 c_{11} + c_{33} + 2 c_{12} + 4 c_{13}) &> 0. 
  \end{split}
\end{equation}
The elastic constants in Table~\ref{tab:pt2si_elastic} satisfy all of the
conditions in Eq.~(\ref{eq:pt2si_stability}).  In particular, $c_{12}$ is
smaller than $c_{11}$ and $c_{13}$ is smaller than the average of $c_{11}$ and
$c_{33}$.  The stability conditions again lead to restrictions on the
magnitude of $B_0$.  We first rewrite $B_0$ as
\begin{equation}
  \label{eq:pt2si_b0_alt}
  B_0 = {\frac{1}{9}} \left[ 6 c_{11} + 3 c_{33} - 2 (c_{11} - c_{12})
      - 2(c_{11} + c_{33} - 2 c_{13}) \right].
\end{equation}
Using Eq.~(\ref{eq:pt2si_b0_alt}) and the first two inequalities in
Eq.~(\ref{eq:pt2si_stability}), we obtain the following result
\begin{equation}
  \label{eq:pt2si_b0_stability1}
  B_0 < {\frac{1}{3}} (2 c_{11} + c_{33}),
\end{equation}
that the bulk modulus must be smaller than the weighted average of $c_{11}$
and $c_{33}$.  Similarly, by substituting instead for $c_{11}$ and $c_{33}$ we
obtain
\begin{equation}
  \label{eq:pt2si_b0_stability2}
  B_0 > {\frac{1}{3}} (c_{12} + 2 c_{13}),
\end{equation}
that the bulk modulus must be larger than the weighted average of $c_{12}$
and $c_{13}$.

The stability restrictions do not tell us anything further about the relative
magnitudes of the various elastic constants.  For example, we find a small
value of $c_{44}$ in comparison to $c_{66}$ which means that the tetragonal
unit cell is more easily deformed by a pure shear about the ${\mathbf a}$- or
${\mathbf b}$-axis in comparison to the ${\mathbf c}$-axis.  We also find that
overall the elastic constants of $\alpha$-Pt$_2$Si are much closer to those of
pure Pt than pure Si.  In particular $c_{11}$ and $c_{33}$ are similar in
magnitude to $c_{11}$ in Pt, but all of these constants are approximately
twice the value of $c_{11}$ in Si.  Similarly $c_{12}$ has approximately the
same magnitude for both $\alpha$-Pt$_2$Si and Pt, although $c_{13}$ in the
silicide is about 30\% smaller.  However, $c_{12}$ in Si is a factor of four
smaller.  Conversely, the value of $c_{44}$ is similar in magnitude for all
three materials, with $c_{66}$ in the silicide being a factor of two larger.
The bulk modulus in the silicide is about 20\% smaller than in Pt but still
more than a factor of two larger than in Si.  The connection between the
magnitudes of the various elastic constants and the chemical bonding has been
explored in detail in a separate study.\cite{Kle00}

In addition to the relaxed elastic constants we also obtained the values of
the dimensionless parameters $\zeta_{i}^{(I)}$ which determine the magnitudes
of the Pt internal displacements themselves,
\begin{equation}
  \label{eq:pt2si_internal_disp}
  \Delta u_{i}^{(I)} = a \zeta_{i}^{(I)} \gamma,
\end{equation}
where $a$ is the lattice constant, $I = 1,4,6$ corresponds to the strains with
symmetry-allowed internal displacements, and $i$ is the Cartesian index ($i =
2,3$ for $I = 6$ and $i = 3$ for $I = 1, 4$).  In principle there could be
contributions to the $\Delta u_{i}^{(I)}$ which are of higher order in
$\gamma$, but since we are only considering the second order elastic constants
the strain energy is only expanded to second order in $\gamma$ [see
Eqs.~(\ref{eq:strain_expansion}) and (\ref{eq:voigt_expansion})], or
equivalently, to second order in the total displacements, $u_{i} + \Delta
u_{i}$.  Thus we need only consider the linear term in
Eq.~(\ref{eq:pt2si_internal_disp}) in the present context.  The calculated
values of the $\zeta_{i}^{(I)}$ parameters are listed in
Table~\ref{tab:internal_disp_params} along with the displacement parameter
associated with $c_{44}$ in Si.  We note that although displacements are
allowed along both the $[010]$ and $[001]$ directions in the case of the
$\alpha$-Pt$_2$Si strain 6 (corresponding to $c_{44}$), the displacements
along $[001]$ are found to be zero to linear order in
$\gamma$.\cite{rem:pt2si_c44_zeta_3} As indicated above, we expect the
internal displacements for strains 1 and 4 to exhibit a symmetry-required
relationship and in keeping with this expectation we find that
$\zeta_{3}^{(I=1)}$ is almost exactly three times larger than
$\zeta_{3}^{(I=4)}$, the difference likely being due to numerical uncertainty.

\subsection{PtSi}
\label{sec:ptsi_elastic}

Nine independent strains are necessary to compute the elastic constants of
orthorhombic PtSi.  We first performed calculations of the elastic constants
with the internal structural parameters $u_{\text{Pt/Si}}$ and
$v_{\text{Pt/Si}}$ held ``frozen'' at their self-consistent equilibrium
values.  These results are listed in the second column of
Table~\ref{tab:ptsi_elastic}.  The $E(\gamma)$ curves are well-fitted by
third-order polynomials in $\gamma$, as can be seen from the small standard
errors in the calculated $c_{ij}$.  The value of $B_0$ obtained from the
elastic constants, $B_{0}=\frac{1}{9}(c_{11} + c_{22} + c_{33} + 2 c_{12} + 2
c_{13} + 2 c_{23})$, agrees reasonably well with the one which was determined
from the calculation of the lattice constants.  This is not surprising since
the lattice constant calculations were also performed with frozen atomic
degrees of freedom.

As expected, the equilibrium atomic positions are not independent of the shape
and size of the unit cell---similarly to the case of $c_{44}$ in Si as well as
$c_{11}$, $c_{12}$, and $c_{44}$ in $\alpha$-Pt$_2$Si, as discussed above.
The relaxed elastic constants for PtSi are listed in the last column of
Table~\ref{tab:ptsi_elastic}.  For example, $c_{44}$ drops from 141.3~GPa to
100.1~GPa when all of the atoms are relaxed.  We find that the Si atoms adjust
to this shear about the ${\mathbf a}$-axis by moving mainly along the
${\mathbf b}$-axis.  Most of the other elastic constants decrease by 10--20\%,
except $c_{23}$ which increases by 8\% and $c_{12}$ which remains
approximately unchanged.  In analogy with the case of $c_{12}$ for tetragonal
$\alpha$-Pt$_2$Si, we note that for an arbitrary strain $c_{12}$, $c_{13}$,
and $c_{23}$ in orthorhombic PtSi never appear isolated but always occur in
combination with other elastic constants in the expression for the
second-order change in the total energy [Eq.~(\ref{eq:voigt_expansion})].
These particular elastic constants are therefore not required to decrease when
relaxation is included, even when the energy is lowered.  Conversely, the
remaining six elastic constants \emph{are} required to decrease when
relaxation lowers the energy because strains can be constructed for which each
appears as the isolated coefficient of the only contribution to the
second-order change in the energy.  In the case of PtSi, the additional
relaxation of the internal degrees of freedom leads to a significant softening
of the elastic constants which must also be taken into account in determining
the bulk modulus.  Therefore, we predict the bulk modulus of PtSi to be
198~GPa, which is 6\% lower than the value of 210~GPa determined in our
frozen-configuration total-energy minimization.  Although we have calculated
the changes in the elastic constants when the internal atomic degrees of
freedom are allowed to relax, we have not explicitly extracted the
corresponding internal displacement parameters as we did for pure Si and
$\alpha$-Pt$_2$Si.

Mechanical stability leads to restrictions on the elastic constants, which for
orthorhombic crystals are as follows\cite{Wal72}
\begin{equation}
  \label{eq:ptsi_stability}
  \begin{split}
    (c_{11} + c_{22} - 2 c_{12}) > 0, \quad
    (c_{11} + c_{33} - 2 c_{13}) &> 0, \\
    (c_{22} + c_{33} - 2 c_{23}) &> 0, \\
    c_{11} > 0, \quad c_{22} > 0, \quad c_{33} &> 0, \\
    c_{44} > 0, \quad c_{55} > 0, \quad c_{66} &> 0, \\
    (c_{11} + c_{22} + c_{33} + 2 c_{12} + 2 c_{13} + 2 c_{23}) &> 0.
  \end{split}
\end{equation}
The elastic constants in Table~\ref{tab:ptsi_elastic} satisfy all of these
conditions and in particular, $c_{12}$ is smaller than the average of $c_{11}$
and $c_{22}$, $c_{13}$ is smaller than the average of $c_{11}$ and $c_{33}$,
and $c_{23}$ is smaller than the average of $c_{22}$ and $c_{33}$.  As in the
case of $\alpha$-Pt$_2$Si, we can obtain restrictions on the magnitude of
$B_0$,
\begin{equation}
  \label{eq:ptsi_b0_stability}
  {\frac{1}{3}} (c_{12} + c_{13} + c_{23}) < B_0 <
  {\frac{1}{3}} (c_{11} + c_{22} + c_{33}),
\end{equation}
that the bulk modulus must be smaller than the average of $c_{11}$, $c_{22}$,
and $c_{33}$ but larger than the average of $c_{12}$, $c_{13}$, and $c_{23}$.

We again find that overall the elastic constants of PtSi are much closer to
those of pure Pt and $\alpha$-Pt$_2$Si as compared to pure Si.  In detail, we
find that $c_{11}$, $c_{22}$, and $c_{33}$ are approximately 10\% smaller on
average in PtSi than in $\alpha$-Pt$_2$Si, and that $B_0$ is 16\% smaller.  In
addition, $c_{12}$, $c_{13}$, and $c_{23}$ for PtSi are close in magnitude to
$c_{13}$ for $\alpha$-Pt$_2$Si, which we saw was about 30\% smaller than
$c_{12}$ in both pure Pt and $\alpha$-Pt$_2$Si.  Finally, $c_{44}$, $c_{55}$,
and $c_{66}$ for PtSi are similar to the values of $c_{44}$ in all three of
the other materials, but still approximately a factor of two smaller than
$c_{66}$ in $\alpha$-Pt$_2$Si.

\subsection{Trends in the elastic constants}
\label{sec:trends_el_const}

\begin{figure}[tbp]
  \centerline{\includegraphics[width=0.8\linewidth]{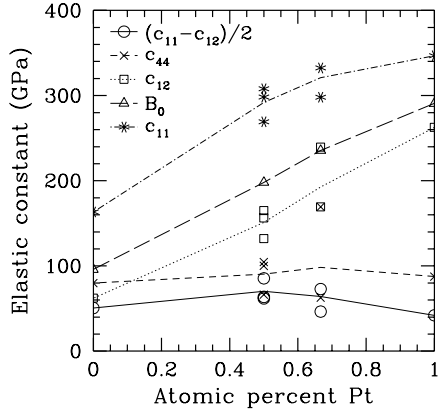}}
  \caption{Trends in the elastic constants as a function of atomic percent Pt
    for pure cubic-diamond-phase Si, orthorhombic PtSi, tetragonal
    $\alpha$-Pt$_2$Si, and fcc Pt.  The different curves correspond to the
    average values of different classes of the individual elastic constants,
    as specified in the legend.  For example, in the case of the dotted-line
    curve labeled as $c_{12}$, the line passes through ${\frac{1}{3}}(c_{12} +
    c_{13} + c_{23})$ in the case of PtSi and through ${\frac{1}{3}}(c_{12} + 2
    c_{13})$ for $\alpha$-Pt$_2$Si ($c_{13} = c_{23}$ for tetragonal
    crystals), while the open squares show the actual values of $c_{12}$,
    $c_{13}$ and $c_{23}$, as appropriate for each material.}
  \label{fig:elas_const_trends}
\end{figure}

The trends of the elastic constants as a function of the atomic percent Pt in
all four materials are plotted in Fig.~\ref{fig:elas_const_trends}.  Each of
the curves corresponds to an average of a different class of elastic
constants, while the symbols show the values of the individual elastic
constants themselves.  As we saw in Eqs.~(\ref{eq:pt2si_b0_stability1}),
(\ref{eq:pt2si_b0_stability2}), and (\ref{eq:ptsi_b0_stability}), mechanical
stability requires that $B_0$ be larger than the average of $c_{11}$,
$c_{22}$, and $c_{33}$ but smaller than the average of $c_{12}$, $c_{13}$, and
$c_{23}$ (note that in the case of $\alpha$-Pt$_2$Si the appropriate averages
are ${\frac{1}{3}}(2 c_{11} + c_{33})$ and ${\frac{1}{3}}(c_{12} + 2 c_{13})$
because $c_{11} = c_{22}$ and $c_{13} = c_{23}$ for tetragonal crystals).
This stability requirement is reflected in the top three curves in
Fig.~\ref{fig:elas_const_trends}.  We also see that these three curves each
increase monotonically from Si to Pt and we note that all three classes of
elastic constants represented by these curves correspond to strains in which
the volume is not fixed.  Conversely, the two lower curves labeled $(c_{11} -
c_{12})/2$ and $c_{44}$ correspond to the two classes of elastic constants in
which the strains are strictly volume-conserving [in the case of PtSi the
lowest solid-line curve and large open circles correspond to elastic constant
combinations ${\frac{1}{4}}(c_{11} + c_{22} - 2 c_{12})$, ${\frac{1}{4}}(c_{11} +
c_{33} - 2 c_{13})$, and ${\frac{1}{4}}(c_{22} + c_{33} - 2 c_{23})$].  We see
that in this case the two sets of averages are approximately constant as a
function of atomic percent Pt.  The significance of this difference in the
trends of volume-conserving versus non-volume-conserving elastic constants is
connected to the nature of the chemical bonding in these materials and has
been addressed in a separate study.\cite{Kle00}

\section{Electronic structure}
\label{sec:electronic_structure}

The self-consistent calculations for the spin-orbit-split energy bands of
$\alpha$-Pt$_2$Si and PtSi were performed using the \textsc{wien97}
implementation\cite{Bla97} of the linear augmented plane wave (LAPW)
method.\cite{And75,Sin94} The local density approximation was used with the
exchange-correlation potential of Perdew and Wang.\cite{pw92} The effects of
the spin-orbit interaction were included in a second-order variational
procedure.\cite{Sin94,Nov97} In the self-consistency cycles approximately 120
irreducible ${\mathbf k}$-points (1000 ${\mathbf k}$-points in the full BZ)
were used in the modified tetrahedron method of Bl\"ochl.\cite{Bloe94} The
energy cutoff used for the plane-wave expansion was \mbox{$k_{\text{max}} =
4.16$ a.u.} resulting in a well converged basis set of about 105 basis
functions per atom.  The experimental values of the lattice constants and
internal structural parameters from Tables~\ref{tab:lattice_constants} and
\ref{tab:internal_coordinates} were used in all cases.  For the purpose of
calculating the density of states (DOS) we again used the tetrahedron method
but with unshifted ${\mathbf k}$-point meshes which included the $\Gamma$
point.  In the case of fcc Pt, cubic-diamond-phase Si, and $\alpha$-Pt$_2$Si
we used a 32$\times$32$\times$32 mesh corresponding to 897 irreducible
${\mathbf k}$-points for the two cubic materials and 2393 for the silicide.
For PtSi we used a 16$\times$24$\times$16 mesh, yielding 1053 irreducible
${\mathbf k}$-points points.

\begin{figure}[tbp]
  \centerline{\includegraphics[width=\linewidth]{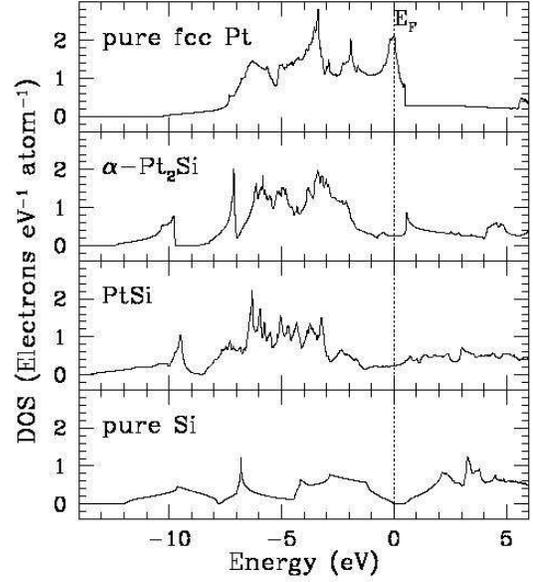}}
  \caption{Total density of states plotted on the same scale for the four
    materials considered in this work, ordered by decreasing Pt content from
    top to bottom.  Only Si is a semiconductor whereas Pt and the two
    silicides are metals.  The valence band maximum in Si is labeled as the
    Fermi level $E_{\text{F}}$.}
  \label{fig:dos_all}
\end{figure}

The total DOS for all four materials is shown on the same scale in
Fig.~\ref{fig:dos_all}.  Although both silicides are metals with a non-zero
DOS at the Fermi level, they are found to be poor metals since the DOS is much
smaller than in the case of pure Pt which is a good metal.  In pure Pt the
Fermi level lies near the top but still within the large DOS features of the
$d-$band but in both silicides the Fermi level lies above these large peaks.
In addition, both silicides exhibit a peak in the DOS at around $-$10~eV which
arises from the Si $s$-orbitals.  We note that the basic features of the
electronic structure, as reflected in the total DOS, do not appear to differ
very much between the two silicides.  The origin of the various features in
the PtSi DOS has been discussed in detail by Franco \emph{et al.}\cite{Fra00}

\begin{figure}[tbp]
  \centerline{\includegraphics[width=\linewidth]{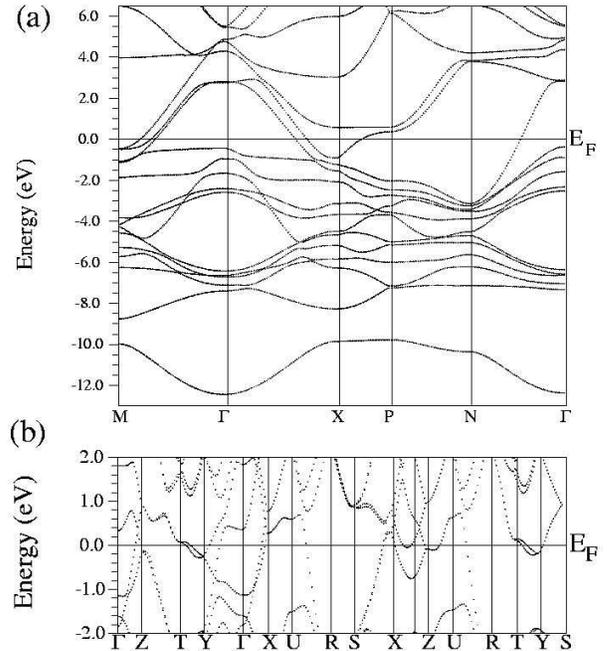}}
  \caption{Spin-orbit split energy bands near the Fermi level for (a)
    $\alpha$-Pt$_2$Si and (b) PtSi.  The primary feature of interest in both
    plots is the relatively large number of low-energy splittings between
    various bands near the Fermi level.}
  \label{fig:so_bands}
\end{figure}

We have also calculated the spin-orbit-split energy bands near the Fermi level
for the two silicides, as shown in Fig.~\ref{fig:so_bands}.  In both cases we
see that there are a sizeable number of small-energy splittings between
different bands throughout the full BZ.  These small splittings are of direct
interest with regard to low-energy inter-band transitions which contribute to
the optical absorption.  In a typical good metal such as Pt, the optical
absorption at low energies is dominated by the free-electron-like Drude
contribution.  However, we saw from Fig.~\ref{fig:dos_all} that in the case of
the silicides they have a low DOS at the Fermi level and consequently are poor
metals.  In this circumstance the Drude contribution will be greatly reduced
and therefore the presence of many low-energy splittings in the bands near the
Fermi level may result in an inter-band contribution to the optical absorption
which is significant even at low energies in the infrared range.

\section{Summary}
\label{sec:summary}

We have carried out an extensive first principles study of two
room-temperature stable Pt silicides, tetragonal $\alpha$-Pt$_2$Si and
orthorhombic PtSi.  We have determined the theoretical equilibrium structural
parameters and cohesive energies for both silicides, as well as pure fcc Pt
and pure cubic-diamond-phase Si.  In particular, we have carried out a large
number of calculations in order to minimize the total energy with respect to
the two lattice constants in tetragonal $\alpha$-Pt$_2$Si and the three
lattice constants and four internal structural parameters of orthorhombic
PtSi.  Our calculated structural parameters for all four materials are in good
agreement with experimental data, validating the method we have used.

A major portion of our effort here has been directed at the elastic constants
in the two silicides.  All of the independent, non-zero elastic constants (6
for $\alpha$-Pt$_2$Si, 9 for PtSi, and 3 each for the two cubic materials)
have been calculated from first principles.  The silicide calculations
required extensive relaxation of the internal degrees of freedom, especially
in the case of the low symmetry structure of orthorhombic PtSi.  Comparing the
elastic constants obtained with and without relaxation we find that relaxation
induces significant changes in the magnitudes of many of the elastic
constants.  In addition, we have explicitly determined the dimensionless
internal displacement parameters for the three strains in $\alpha$-Pt$_2$Si
for which they are non-zero.  We also note that the value of $c_{44}$ in pure
fcc Pt was found to be extremely sensitive to the number of ${\mathbf
k}$-points, much more so than any of the other elastic constants we
calculated.  This sensitivity results from the large number of van Hove
singularities close to the Fermi level.

We have investigated the trends in the calculated elastic constants, both the
trends within a given material as well as between materials.  The requirement
of mechanical stability places specific restrictions on the relative
magnitudes of some of the elastic constants within a given material,
including, for example, a restriction on the bulk modulus that $B_0 <
\frac{1}{3}(c_{11} + c_{22} + c_{33})$ and $B_0 > \frac{1}{3}(c_{12} +
c_{13} + c_{23})$.  With regard to the trends among the four materials, we
find that in the metals the elastic constant expressions which correspond to
volume-conserving strains are always smaller than those which correspond to
strains which do not conserve volume.  This also turns out to be true in Si
with the exception of $c_{12}$ which is less than $c_{44}$.  However, the
difference in magnitudes between volume-conserving and non-volume-conserving
elastic constants is largest on average in Pt and gets smaller in the
progression Pt $\rightarrow$ $\alpha$-Pt$_2$Si $\rightarrow$ PtSi
$\rightarrow$ Si.  In general, the volume-conserving elastic constants have
similar magnitudes in all four materials while the non-volume-conserving
elastic constants follow this same progression.  In particular, the bulk
modulus is found to be a very nearly linear function of the atomic percentage
of Pt.  Klepeis \emph{et al.}\cite{Kle00} have studied the close connection
between the various trends in the elastic constants and the chemical bonding
in the Pt silicides.

The calculated electronic structure demonstrates that the two silicides are
poor metals with a low density of states at the Fermi level, and consequently
we expect that the Drude component of the optical absorption should be much
smaller than in good metals such as pure Pt.  In addition, we find a large
number of small-energy differences between various bands near the Fermi level
in the calculated spin-orbit-split band structure for the two silicides.
These two circumstances suggest that it may be important to include the
interband contribution to the optical absorption as well, even in the infrared
region.

\acknowledgments

This work was performed in part under the auspices of the U.~S. Department of
Energy, Office of Basic Energy Sciences, Division of Materials Science by the
University of California Lawrence Livermore National Laboratory under contract
No.~W--7405--Eng--48.  Partial support was also provided by Deutsche
Forschungsgemeinschaft, SFB 292 ``Multicomponent Layered Systems.''

\newpage

\widetext

\newpage

\begin{table}[tbp]
  \caption{Structural characterization of the materials studied in this work.
    The column labeled ``Structure'' gives the Strukturbericht designations.
    The columns ``Space Group'' (name and number), ``Site'' (multiplicity and
    Wyckoff letter), and ``1st position'' follow
    Ref.~\protect\onlinecite{ITCA}.  See Table~\ref{tab:internal_coordinates}
    for the values of the internal structural parameters $u$ and $v$.}
  \renewcommand{\arraystretch}{1.5}
  \begin{tabular}{llccllcc}
    Material & Structure  & \multicolumn{2}{c}{Space group}  &
    \multicolumn{2}{c}{Site} & 1st position & Ref.\\
    \hline
    Pt                  & $A1$      & $Fm\bar{3}m$ & 225
        & Pt & $4a$ & $0,0,0$ & \onlinecite{Pea64} \\
    $\alpha$-Pt$_2$Si   & $L'2_{b}$ & $I4/mmm$     & 139
        & Pt & $4d$ & $0,\frac{1}{2},\frac{1}{4}$ & \onlinecite{Ram78} \\
                        &           &              &
        & Si & $2a$ & $0,0,0$ & \\
    PtSi                & $B31$     & $Pnma$       & 62
        & Pt & $4c$ & $u_{\text{Pt}},\frac{1}{4},v_{\text{Pt}}$
                      & \onlinecite{Grae73,Vil91}\\
                        &           &              &
        & Si & $4c$ & $u_{\text{Si}},\frac{1}{4},v_{\text{Si}}$ &  \\
    Si                  & $A4$      & $Fd\bar{3}m$ & 227
        & Si & $8a$ & $\frac{1}{8},\frac{1}{8},\frac{1}{8}$
                      & \onlinecite{LBnsIII/17a}
  \end{tabular}
  \label{tab:crystal_structures}
\end{table}

\begin{table}[tbp]
  \caption{Theoretical and experimental lattice constants (in a.u.) and bulk
    moduli $B_{0}$ (in GPa).  The theoretical $B_{0}$ were determined from a
    fit to a Murnaghan equation of state for Pt and Si, and to a four-term
    Birch-Murnaghan equation of state for $\alpha$-Pt$_2$Si and PtSi.  $B_0$
    listed here for PtSi was calculated for fixed values of the internal
    structural parameters (see also Table~\ref{tab:ptsi_elastic}).  The
    experimental value of the bulk modulus for Pt is extrapolated to 0~K,
    whereas all of the other experimental numbers are given for room
    temperature.}
  \begin{tabular}{llddddc}
    Material &        & $a_{0}$ & $b_{0}$ & $c_{0}$ & $B_{0}$
      & Ref.\\
    \hline
    Pt       & theor. &   7.403 &         &         & 287.8 \\
             & exp.   &   7.415 &         &         & 288.4
      & \onlinecite{LBnsIII/6,Mac65} \\
    $\alpha$-Pt$_2$Si
             & theor. &   7.407 &         &  11.241 & 233.5 \\
             & exp.   &   7.461 &         &  11.268 &  ---
      & \onlinecite{Ram78} \\
    PtSi     & theor. &  10.583 &   6.774 &  11.195 & 210.0 \\
             & exp.   &  10.539 &   6.778 &  11.180 &  ---
      & \onlinecite{Grae73} \\
    Si       & theor. &  10.22  &         &         &  95.9 \\
             & exp.   &  10.26  &         &         &  98.8
      & \onlinecite{LBnsIII/29a}
  \end{tabular}
  \label{tab:lattice_constants}
\end{table}

\begin{table}[tbp]
  \caption{Calculated and experimental internal structural parameters of PtSi.
    The parameters $u$ and $v$ are the same as those specified in
    Table~\ref{tab:crystal_structures}.  These parameters were calculated both
    for the experimental lattice constants as well as the self-consistent
    theoretical lattice constants (Table~\ref{tab:lattice_constants}).  The
    experimental internal parameters for PtSi are also listed.}
  \begin{tabular}{ldddd}
    PtSi & $u_{\text{Pt}}$ & $v_{\text{Pt}}$ & $u_{\text{Si}}$
      & $v_{\text{Si}}$ \\
    \hline
    At expt. lattice constants  & 0.9981 & 0.1915 & 0.1777 & 0.5845 \\
    At theor. lattice constants & 0.9977 & 0.1919 & 0.1782 & 0.5841 \\
    Experiment\cite{Grae73}  & 0.9956 & 0.1922 & 0.177  & 0.583  \\
  \end{tabular}
  \label{tab:internal_coordinates}
\end{table}

\begin{table}[tbp]
  \caption{Parameters describing the basis used in the FPLMTO calculations.
    $R_{\text{mt}}$ is the muffin-tin radius in a.u., $L_{\text{max}}$ is the
    upper limit on the angular momentum expansion of the smoothed Hankel
    functions about a given atomic site, $K_{\text{max}}$ is the order of the
    biorthogonal polynomials used in this expansion, $R_{\text{sm}}$ is the
    smoothing radius in a.u., and $-\kappa^2$ is the decay energy.  The total
    number of basis functions per atom is 17 for Pt and 13 for Si.  See
    Ref.~\protect\onlinecite{nfp97} for a more complete description of the
    parameters.}
  \begin{tabular}{ldddcdd}
    Atom & $R_{\text{mt}}$ & $L_{\text{max}}$ & $K_{\text{max}}$
     & \multicolumn{3}{c}{Basis} \\
         &     &   &   & $L$\,-Block & $R_{\text{sm}}$ & $-\kappa^2$\\
    \hline
    Pt   & 2.2 & 3 & 4 &     $s$     &      2.865      &  $-$1.02  \\
         &     &   &   &     $p$     &      2.130      &  $-$1.08  \\
         &     &   &   &     $p$     &      1.302      &  $-$1.53  \\
         &     &   &   &     $d$     &      1.000      &  $-$0.89  \\
         &     &   &   &     $d$     &      2.123      &  $-$0.42  \\
    Si   & 2.1 & 3 & 5 &     $s$     &      1.908      &  $-$1.296 \\
         &     &   &   &     $p$     &      1.627      &  $-$0.302 \\
         &     &   &   &     $d$     &      1.601      &  $-$1.496 \\
         &     &   &   &   $s$,$p$   &      2.200      &  $-$2.000
  \end{tabular}
  \label{tab:nfp_basis}
\end{table}

\begin{table}[tbp]
  \caption{Cohesive energies $E_{\text{coh}}$ and heats of formation $\Delta
    H_{\text{f}}$ in eV/atom.  The experimental standard heat of formation is
    given for $T=298.15$~K whereas theoretical values are valid for $0$~K and
    do not contain any corrections for zero-point vibrations.  The
    experimental cohesive energies for Pt and Si also correspond to $0$~K.}
  \begin{tabular}{llddc}
    Material &        & $E_{\text{coh}}$ & $\Delta H_{\text{f}}$ & Ref. \\
    \hline
    Pt       & theor. &       7.27       &       0      &      \\
             & exp.   &       5.84       &       0      & \onlinecite{Kit86} \\
    $\alpha$-Pt$_2$Si
             & theor. &       7.24       &    $-$0.65   &      \\
             & exp.   &       6.08       &    $-$0.64   &
      \onlinecite{Top86,rem:Ecoh_Pt2Si} \\
    PtSi     & theor. &       6.93       &    $-$0.67   &      \\
             & exp.   &       5.85       &    $-$0.62   &
      \onlinecite{Cha93,rem:Ecoh_PtSi} \\
    Si       & theor. &       5.23       &       0      &      \\
             & exp.   &       4.63       &       0      & \onlinecite{Kit86}
  \end{tabular}
  \label{tab:cohesive_energies}
\end{table}

\begin{table}[tbp]
  \caption{Parameterizations of the three strains used to calculate the three
    elastic constants of cubic Pt and Si (also used in
    Ref.~\protect\onlinecite{Meh90}).  The energy expressions were 
    obtained from Eq.~(\protect\ref{eq:voigt_expansion}).  Strains $I=1$ and
    $I=3$ are strictly volume-conserving to all orders in the strain parameter
    $\gamma$.  If we restrict ourselves to linear order only then
    $e^{(1)}_{3}=-2\gamma$ and $e^{(3)}_{3} = 0$, with volume conservation
    preserved to linear order as well.}
  \renewcommand{\arraystretch}{1.5}
  \begin{tabular}{ccc}
    Strain $I$
      & Parameters (unlisted $e_{i}=0$)
      & \text{$\Delta E/V$ to $O(\gamma^{2})$} \\
    \hline
    1 & $e_{1}=e_{2}=\gamma,\ e_{3}=(1+\gamma)^{-2}-1$
      & $3(c_{11} - c_{12})\gamma^{2}$ \\
    2 & $e_{1}=e_{2}=e_{3}=\gamma$
      & $\frac{3}{2}(c_{11}+2 c_{12})\gamma^{2}$ \\
    3 & $e_{6}=\gamma,\ e_{3}=\gamma^{2}(4-\gamma^{2})^{-1}$
      & $\frac{1}{2}c_{44}\gamma^{2}$
  \end{tabular}
  \label{tab:cubic_strains}
\end{table}

\begin{table}[tbp]
  \caption{Parameterizations of the six strains used to calculate the six
    elastic constants of tetragonal $\alpha$-Pt$_2$Si (taken from 
    Ref.~\protect\onlinecite{Alo91}).  The energy expressions 
    were obtained from Eq.~(\protect\ref{eq:voigt_expansion}).}
  \renewcommand{\arraystretch}{1.5}
  \begin{tabular}{ccc}
    Strain $I$
      & Parameters (unlisted $e_{i}=0$)
      & \text{$\Delta E/V$ to $O(\gamma^{2})$} \\
    \hline
    1 & $e_{1}=2\gamma,\ e_{2}=e_{3}=-\gamma$
      & $\frac{1}{2}(5 c_{11} - 4 c_{12} - 2 c_{13} + c_{33})\gamma^{2}$\\
    2 & $e_{1}=e_{2}=-\gamma,\ e_{3}=2\gamma$
      & $(c_{11} + c_{12} - 4 c_{13} + 2 c_{33})\gamma^{2}$\\
    3 & $e_{1}=e_{2}=\gamma,\ e_{3}=-2\gamma,\ e_{6}=2\gamma$
      & $(c_{11} + c_{12} - 4 c_{13} + 2 c_{33} + 2 c_{66})\gamma^{2}$\\
    4 & $e_{1}=\gamma$
      & $\frac{1}{2}c_{11}\gamma^{2}$\\
    5 & $e_{3}=\gamma$
      & $\frac{1}{2}c_{33}\gamma^{2}$\\
    6 & $e_{4}=2\gamma$
      & $2 c_{44}\gamma^{2}$
  \end{tabular}
  \label{tab:pt2si_strains}
\end{table}

\begin{table}[tbp]
  \caption{Parameterizations of the nine strains used to calculate the nine
    elastic constants of orthorhombic PtSi.  The energy expressions were
    obtained from Eq.~(\protect\ref{eq:voigt_expansion}).}
  \renewcommand{\arraystretch}{1.5}
  \begin{tabular}{ccc}
    Strain $I$
      & Parameters (unlisted $e_{i}=0$)
      & \text{$\Delta E/V$ to $O(\gamma^{2})$} \\
    \hline
    1 & $e_{1}=\gamma$
      & $\frac{1}{2}c_{11}\gamma^{2}$\\
    2 & $e_{2}=\gamma$
      & $\frac{1}{2}c_{22}\gamma^{2}$\\
    3 & $e_{3}=\gamma$
      & $\frac{1}{2}c_{33}\gamma^{2}$\\
    4 & $e_{1}=2\gamma,\ e_{2}=-\gamma,\ e_{3}=-\gamma$
      & $\frac{1}{2}(4 c_{11} - 4 c_{12}
                     - 4 c_{13} + c_{22} + 2 c_{23} + c_{33}) \gamma^{2}$\\
    5 & $e_{1}=-\gamma,\ e_{2}=2\gamma,\ e_{3}=-\gamma$
      & $\frac{1}{2}(  c_{11} - 4 c_{12}
                     + 2 c_{13} + 4 c_{22} - 4 c_{23} + c_{33}) \gamma^{2}$\\
    6 & $e_{1}=-\gamma,\ e_{2}=-\gamma,\ e_{3}=2\gamma$
      & $\frac{1}{2}(  c_{11} + 2 c_{12}
                     - 4 c_{13} + c_{22} - 4 c_{23} + 4 c_{33}) \gamma^{2}$\\
    7 & $e_{4}=\gamma$
      & $\frac{1}{2}c_{44}\gamma^{2}$\\
    8 & $e_{5}=\gamma$
      & $\frac{1}{2}c_{55}\gamma^{2}$\\
    9 & $e_{6}=\gamma$
      & $\frac{1}{2}c_{66}\gamma^{2}$
  \end{tabular}
  \label{tab:ptsi_strains}
\end{table}

\begin{table}[tbp]
  \caption{Elastic constants of Pt and Si.  Calculations were carried out at
    the theoretical self-consistent lattice constants of $a_{\text{Pt}} =
    7.403$ a.u. and $a_{\text{Si}} = 10.22$ a.u.  The theoretical value of
    $c_{44}$ in parentheses for Si is the ``frozen'' value obtained without
    allowing for internal relaxation.  The bulk modulus is calculated from the
    elastic constants as $B_{0}=\frac{1}{3}(c_{11} + 2 c_{12})$.  In
    parentheses we give $B_{0}$ from the fit to a Murnaghan equation of state.
    Experimental values are extrapolated to 0~K.  All values are in units of
    GPa.}
  \begin{tabular}{l*{4}{r@{}l}}
      & \multicolumn{2}{c}{Pt Theory}
      & \multicolumn{2}{c}{Pt Expt.\protect\cite{Mac65}}
      & \multicolumn{2}{c}{Si Theory}
      & \multicolumn{2}{c}{Si Expt.\protect\cite{LBnsIII/29a}} \\
    \hline
    $c_{11}$ & 346&.8$\pm$0.5           & 358&
             & 163&.45$\pm$0.03         & 165&   \\
    $c_{12}$ & 262&.7$\pm$0.3           & 254&
             &  62&.13$\pm$0.02         &  63&   \\
    $c_{44}$ &  87&.5$\pm$0.3           &  77&
             &  79&.85$\pm$0.02 (108.6) &  79&.1 \\
    $B_{0}$  & 290&.8$\pm$0.3 (287.8)   & 288&.4
             &  95&.90$\pm$0.02 (95.9)  &  97&.0 \\
  \end{tabular}
  \label{tab:pt+si_elastic}
\end{table}

\begin{table}[tbp]
  \caption{Elastic constants of $\alpha$-Pt$_2$Si.  Calculations were
    performed at the theoretical self-consistent lattice constants
    (Table~\ref{tab:lattice_constants}).  ``Frozen'' refers to fixed atomic
    positions, whereas ``relaxed'' indicates that a relaxation of the atomic
    positions was carried out.  Parentheses denote values where no internal
    relaxation was necessary because of symmetry constraints (small variations
    in these values come from using a slightly more stringent convergence
    criterion on the energy).  The bulk modulus is calculated from the elastic
    constants as $B_{0}=\frac{1}{9}(2 c_{11} + c_{33} + 2 c_{12} + 4 c_{13})$.
    $B_{0}^{\text{Birch}}$ is from a Birch-Murnaghan fit.  No experimental
    data is available.  All values are in units of GPa.}
  \begin{tabular}{l*{2}{r@{}l}}
    $\alpha$-Pt$_2$Si
      & \multicolumn{2}{c}{frozen}
      & \multicolumn{2}{c}{relaxed} \\
    \hline
    $c_{11}$ & 347&.2$\pm$1.2 &  332&.4$\pm$0.9  \\
    $c_{33}$ & 297&.5$\pm$0.5 & (298&.0$\pm$0.4) \\
    $c_{12}$ & 225&.0$\pm$1.2 &  239&.6$\pm$1.0  \\
    $c_{13}$ & 169&.3$\pm$0.9 & (169&.4$\pm$0.8) \\
    $c_{44}$ &  75&.4$\pm$0.3 &   62&.7$\pm$0.5  \\
    $c_{66}$ & 169&.5$\pm$5.2 & (169&.3$\pm$5.2) \\
    $B_{0}$  & 235&.4$\pm$0.6 & (235&.5$\pm$0.5) \\
    $B_{0}^{\text{Birch}}$ & 233&.5 & ---
  \end{tabular}
  \label{tab:pt2si_elastic}
\end{table}

\begin{table}[tbp]
  \caption{Dimensionless internal displacement parameters $\zeta$ for Si
    [Eq.~(\protect\ref{eq:si_internal_disp}), strain 3 from
    Table~\protect\ref{tab:cubic_strains}] and $\alpha$-Pt$_2$Si
    [Eq.~(\protect\ref{eq:pt2si_internal_disp}), strains 1, 4, and 6 from
    Table~\protect\ref{tab:pt2si_strains}].  In the
    case of strain 6 for $\alpha$-Pt$_2$Si displacements were calculated along
    both the [010] and [001] directions, whereas displacements only along the
    [001] direction are allowed by symmetry for strains 1 and 4.  In addition,
    there is a strict symmetry-required relationship between the displacements
    in strains 1 and 4 (see text).  Experimental data is available only for
    Si.}
  \begin{tabular}{lcddc}
    Material          & Strain $I$  &  Theory  & Experiment & Ref. \\
    \hline
    Si                &     3       &    0.53  & 0.54 & \onlinecite{Cou87} \\
    $\alpha$-Pt$_2$Si &     1       &    0.22  &  -- \\
                      &     4       &    0.074 &  -- \\
                      &     6 [010] & $-$0.12  &  -- \\
                      &     6 [001] &    0.00  &  -- \\
  \end{tabular}
  \label{tab:internal_disp_params}
\end{table}

\begin{table}[tbp]
  \caption{Elastic constants of PtSi calculated at the theoretical
    self-consistent lattice constants (Table~\ref{tab:lattice_constants}).
    The second column shows the elastic constants obtained when the internal
    structural parameters were held fixed at their theoretical self-consistent
    values (Table~\ref{tab:internal_coordinates}).  In the third column these
    were allowed to relax.  The bulk modulus is calculated from the elastic
    constants as $B_{0}=\frac{1}{9}(c_{11} + c_{22} + c_{33} + 2 c_{12} + 2
    c_{13} + 2 c_{23})$.  $B_{0}^{\text{Birch}}$ is from a Birch-Murnaghan fit
    obtained with frozen values of the internal structural parameters.  No
    experimental data is available.  All values are in units of GPa.}
  \begin{tabular}{l*{2}{r@{}l}}
    PtSi
      & \multicolumn{2}{c}{frozen}
      & \multicolumn{2}{c}{relaxed}\\
    \hline
    $c_{11}$ & 327&.5$\pm$1.2 & 298&.2$\pm$1.2 \\
    $c_{22}$ & 313&.8$\pm$0.0 & 269&.3$\pm$0.8 \\
    $c_{33}$ & 345&.9$\pm$0.1 & 308&.0$\pm$0.6 \\
    $c_{12}$ & 157&.7$\pm$0.6 & 156&.4$\pm$0.8 \\
    $c_{13}$ & 162&.9$\pm$0.6 & 132&.2$\pm$0.7 \\
    $c_{23}$ & 153&.4$\pm$0.1 & 165&.1$\pm$0.6 \\
    $c_{44}$ & 141&.3$\pm$0.3 & 100&.1$\pm$0.4 \\
    $c_{55}$ & 113&.1$\pm$0.1 & 104&.5$\pm$0.1 \\
    $c_{66}$ &  74&.2$\pm$0.2 &  66&.3$\pm$0.4 \\
    $B_{0}$  & 215&.0$\pm$0.2 & 198&.1$\pm$0.3 \\
    $B_{0}^{\text{Birch}}$ & 210&.0 & ---
  \end{tabular}
  \label{tab:ptsi_elastic}
\end{table}


\begin{references}
\bibitem{rem:nfp_basis}
The basis used here is ``minimal'' in the sense that removal of any one of the
  $L$\,-blocks listed in Table~\ref{tab:nfp_basis} would result in a
  significant reduction in accuracy.  However, we have tested this basis on
  the equilibrium lattice constants and elastic constants of pure Pt and pure
  Si (see Tables~\ref{tab:lattice_constants} and \ref{tab:pt+si_elastic}).
  For example, adding a third $d$\,-block for Pt results in only a 0.2\%
  reduction in the theoretical equilibrium lattice constant.
\bibitem{rem:geomSC}
Alternatively, one could take geometry relaxations into account while
  simultaneously varying the lattice constants, but this approach is
  computationally demanding.  Nevertheless, this method was required for the
  elastic constant calculations discussed in Sec.~\ref{sec:method_elastic}, as
  demonstrated by the considerable differences between ``frozen'' and
  ``relaxed'' elastic constants (see Tables~\ref{tab:pt2si_elastic} and
  \ref{tab:ptsi_elastic}).
\bibitem{rem:pt2si_strains}
We could also have used a simpler strain 3 in Table~\ref{tab:pt2si_strains} in
  order to determine $c_{66}$, namely $e_{6}=2\gamma$ with $\Delta E/V = 2
  c_{66} \gamma^{2}$.
\bibitem{rem:pt2si_c44_zeta_3}
The calculated displacements along $[001]$ for strain 6 are found to be
  approximately a factor of 30 smaller than the displacements along $[010]$
  and they do not depend on the sign but only the magnitude of $\gamma$.
\bibitem{rem:Ecoh_Pt2Si}
The ``experimental'' cohesive energy for $\alpha$-Pt$_2$Si was calculated as
  $E_{\text{coh}} = {\frac{2}{3}} E_{\text{coh}}(\text{Pt}) + {\frac{1}{3}}
  E_{\text{coh}}(\text{Si}) - \Delta
  H^{298.15}_{\text{f}}(\alpha\text{-Pt}_2\text{Si})$ from the experimental
  cohesive energies\cite{Kit86} and the molar heat of formation,\cite{Top86}
  without correcting for the difference between 298.15 and 0~K.
\bibitem{rem:Ecoh_PtSi}
The ``experimental'' cohesive energy for PtSi was calculated as
  $E_{\text{coh}} = {\frac{1}{2}} E_{\text{coh}}(\text{Pt}) + {\frac{1}{2}}
  E_{\text{coh}}(\text{Si}) - \Delta H^{298.15}_{\text{f}}(\text{PtSi})$ from
  the experimental cohesive energies\cite{Kit86} and the molar heat of
  formation,\cite{Cha93} without correcting for the difference between 298.15
  and 0~K.
\end{references}
\end{document}